\documentclass[%
 reprint,
groupedaddress,
unsortedaddress,
 amsmath,amssymb,
 aps,
prb,
]{revtex4-2}
\usepackage{graphicx, multirow}
\usepackage{dcolumn}
\usepackage{bm}
\usepackage[colorlinks=true,allcolors=blue]{hyperref}
\usepackage{lipsum}
\usepackage[mathlines]{lineno}
\usepackage{tabularx}
\usepackage{xcolor}
\usepackage{tabu}
\usepackage{booktabs}
\usepackage{array}

\begin{document}
	

\title{Magnetotransport Properties and Fermi Surface Topology of Nodal line Semimetal InBi }

\author{Sambhab Dan$^{a}$, Kuldeep Kargeti$^{b}$, R. C. Sahoo$^{c}$, Shovan Dan$^{d}$, Debarati Pal$^{a}$, Sunil Verma$^{a}$, Sujay Chakravarty$^{e}$, S. K. Panda$^{b}$, S Patil$^{a}$}

\email{spatil.phy@iitbhu.ac.in}

\affiliation{$^{a}$Department of Physics, Indian Institute of Technology (BHU), Varanasi 221005, India}
\affiliation{$^{b}$Department of Physics, Bennett University, Greater Noida 201310, Uttar Pradesh, India}
\affiliation {$^{c}$Department of Chemical Science and Engineering, Tokyo Institute of Technology, 2-12-1 Ookayama, Meguro, Tokyo 152-8552, Japan}
\affiliation{$^{d}$Department of Condensed Matter Physics and Materials Science, Tata Institute of Fundamental Research, Homi Bhabha Road, Colaba, Mumbai 400005, India}
\affiliation{$^{e}$UGC-DAE Consortium for Scientific Research, Kalpakkam Node, Kokilamedu 603104, India}



\date{\today}
\begin{abstract}
	 In the present study, we have discussed the up-turn behavior in the resistivity pattern of the topological nodal line semimetal InBi. We argued that such nature could be generalized with a mathematical model, that can be applied to any compounds exhibiting similar behavior. The extremely high magnetoresistance (XMR) has also been explained by the carrier compensation in the compound, estimated from the Hall conductivity. Moreover, from the study of Subhnikov-de Haas (SdH) oscillation and density functional theory (DFT), we obtained the complete three-dimensional (3D) Fermi surface topology of the compound InBi. A detailed understanding of carriers' behavior has been discussed using those studies. We have also unfurled the topology of each electron and hole pocket and its possible modulation with electron and hole doping.  
\end{abstract}

\maketitle


\section {Introduction}
Dealing with exotic bandstructure of the topological nodal line semimetals (TNLS) is flourishing field in modern condensed matter physics. The electronic structure originating from the band theory describes the topological classification of the materials. For example,  in the case of a topological insulator (TI), the surface band is topologically non-trivial, whereas the bulk band remains in an insulating phase\cite{moore2010birth}. Unlike TI, a lot of metals and semimetals also poses topologically non-trivial bulk band. Nevertheless, the recently discovered Dirac and Weyl semimetals significantly divert the attention toward topological semimetals\cite{armitage2018weyl}. Unlike the TI, Dirac and Weyl semimetals show the Dirac cone in the bulk band, and conduction occurs through its bulk state. In such a semimetallic phase, the conduction and valence band touches at the Dirac Point (DP), which can be called a zero-dimensional nodal point. But, in the case of TNLS, this nodal point becomes one dimension nodal line extended either in a line or a loop in the momentum space. The compounds having such features in the electronic bandstructure are called either nodal line\cite{burkov2011topological}, nodal chain\cite{bzduvsek2016nodal}, nodal knot\cite{bi2017nodal}, nodal-link\cite{yan2017nodal, chang2017weyl, chen2017topological} or nodal-net\cite{kim2015dirac, yu2015topological}.

Recently discovered a few promising nodal line materials are ZrSiX (X= S, Se, Te )\cite{schoop2016dirac, hosen2017tunability}, Ag$_2$S\cite{huang2017topological}, AX$_2$ (A = Ca, Sr, Ba; X = Si, Ge, Sn).\cite{huang2016topological}, Ca$_3$P$_2$\cite{chan20163}, CaP$_3$\cite{xu2017topological}, CaTe\cite{du10cate}, CaAuAs\cite{singh2018spin}, CaAgX (X = P, As)\cite{takane2018observation}, IrF$_4$\cite{bzduvsek2016nodal}, Mg$_3$Bi$_2$\cite{chang2019realization}, XB$_2$ (X = Ti, Zr)\cite{lou2018experimental}, SrAs$_3$\cite{song2020photoemission}, Co$_2$TiX (Si, Ge or Sn)\cite{chang2016room}, Ta$_3$X (X = Al, Ga, Sn, Pb)\cite{zhang2018nodal}, X$_2$Y (X = Ca, Sr, Ba; Y = As, Sb, Bi)\cite{niu2017two}, YCoC$_2$\cite{xu2019topological}, MnN\cite{wang2019two} and YH$_3$\cite{shao2018nonsymmorphic} \textit{etc}. As one dimensional band-crossing  contains a larger density of states compering to the nodal points; such materials show novel properties like large spin Hall effect\cite{sun2017dirac}, long-range Coulomb interaction\cite{huh2016long}, and flat Landau level\cite{rhim2015landau}. \textit{etc}. Interestingly, when spin-orbit coupling comes into the scenario, nodal line of the material is either protected by inversion, reflection, and nonsymmorphic symmetry or even gaped due to the lack of such symmetries\cite{burkov2011topological, fang2015topological, fang2016topological, kim2015dirac, weng2015topological}. For example, reflection of the Ta atomic plane protects the nodal line in XTaSe$_2$ (X = Pb, Tl)\cite{bian2016drumhead, bian2016topological}. In the case of Cu$_3$PdN, nodal lines are protected by inversion symmetry\cite{yu2015topological}. For ZrSiS\cite{schoop2016dirac}  and InBi\cite{ekahana2018investigation}, nodal lines are protected by non-symmorphic symmetry. Although a vast studies is done on ZrSiS, InBi remains the least studied compound till the date. InBi has two distinct nodal lines protected by non-symmorphic symmetry\cite{ekahana2018investigation} and also having topologically non-trivial type-II Dirac cone protected by four-fold cyclic symmetry\cite{ekahana2018investigation}. The compound InBi exhibits unique transport properties \textit{viz.} extremely high magnetoresistance (XMR)\cite{okawa2018extremely}, highly anisotropic MR\cite{okawa2018extremely}, and pressure-induced superconductivity\cite{tissen1998superconductivity}. The compound also shows  an upturn nature of the temperature (\textit{T}) dependent resistivity ($\rho$)\cite{okawa2018extremely} in the presence of magnetic field (\textbf{B}), but a detailed discussion on that topic is not available to the date. The quantum oscillation study of the compound is also reported earlier\cite{meyer1974haas}, but a deep understanding of Fermi surface topology and its role in transport is still lacking. 

In this rapid communication, we report structural, transport, electronic bandstructure and Fermi surface topology of InBi. The report is organized in the following way. We discussed structural properties of the compound in Sec. \ref{A}. In Sec. \ref{B}, we have studied the magnetotransport properties of InBi. In this section, we have focused on the interesting result, \textit{viz.} up-turn nature in $\rho (T)$. We have given a mathematical model based on up-turn behaviour. Our experimental result well justifies the model. Another aspect of our study is establishing the role of electron and hole carriers in the transport phenomena. Although the attentive studies have been performed earlier\cite{okawa2018extremely}, the important role of carrier concentration in the transport studies was still missing. We bring some exciting results to address the issue. The results are compelling and could unravel the origin of XMR in InBi. In Sec. \ref{C}, We have discussed the Fermi surface topology by Shubnikov-de Haas (SdH) oscillation. We verified our experimental result by theoretical analysis based on density functional theory (DFT). Our DFT result satisfactorily matches the experimental data. Accumulating the empirical and theoretical tools, we have successfully mapped the compound's Fermi surface. Such work gives a complete three dimensional (3D) visualization of the Fermi surface. We provide a detailed discussion of each Fermi sheet (FS). The modulation of each FS due to electron and hole doping and the FS's role in transport phenomena are also discussed rigorously. At the end of this section, we have discussed about some important physical parameters derived from our SdH oscillation.

\section{Experimental details and theoretical methods}

We prepared single crystal InBi using the modified Bridgman method. Elemental In (Alfa Aesar) and Te (Alfa Aesar), with at least 99.9$\%$ purity, have been used as the starting materials. The stoichiometric amount of starting materials was taken in an evacuated (10$^{-6}$ mBar) quartz ampule. The ingots were melted at 200$^{\circ}$ C in the quartz ampule and kept for 24 hr at that temperature. The liquid of the ingots was subsequently cooled to 150$^{\circ}$ C with a cooling rate of 20$^{\circ}$.hr$^{-1}$. Then we cooled the molten material from 150$^{\circ}$ to 30$^{\circ}$ C in slow cooling rate 3$^{\circ}$.hr$^{-1}$. As the bottom edge of the quartz tube is a conical shape, the nucleation process starts from the bottom point of the quartz tube, and the crystallization process has been completed at this stage. The shiny conical shape single crystal is obtained from the quartz tube. The small parts of the crystal have been used for further characterization.

X-ray diffraction (XRD) of powder compound was performed for structural analysis. We also performed XRD on compound’s cleaved surface to check the single crystallinity. The cleaved surface's XRD study identifies the particular planes’ Miller indices. The Hall study of the compound was performed on a rectangle-shaped sample with the four-point probe in the van der Pauw configuration. The Hall study was completed using a 9T Physical Property Measurement System(PPMS), and the Hall voltages were detected along the direction perpendicular to the current flow in the sample plane. The MR measurement was performed in the four probe puck in 16 T PPMS. During MR and Hall measurements, the external magnetic field was applied perpendicular to the \textit{ab}-plane. 

To interpret our experimental results, we have carried out density functional theory (DFT) calculations using full potential linearized augmented plane-wave based method implemented in WIEN2k code\cite{schwarz2003solid}. All the calculations are performed in a non-spin polarized set-up with generalized gradient approximation (GGA)\cite{perdew1996generalized} approach for the exchange-correlation functional. We include spin-orbit coupling (SOC) into the system, and all result is obtained from GGA+SOC calculation. In order to obtain a very accurate Fermi energy, $R_{mt}K_{max}$ value is considered to be 7 and Brillouin zone integration is done using tetrahedron method\cite{blochl1994improved} with a sufficiently dense \textit{k}-mesh ($14\times14\times14$). We have generated the 3D Fermi surface using wannier90\cite{mostofi2014updated} and used xcrysden\cite{kokalj1999xcrysden} for plotting.

\section{Results and discussion}

\subsection{Structural characterization}
\label{A}
\begin{figure}
	\includegraphics[width=0.49\textwidth]{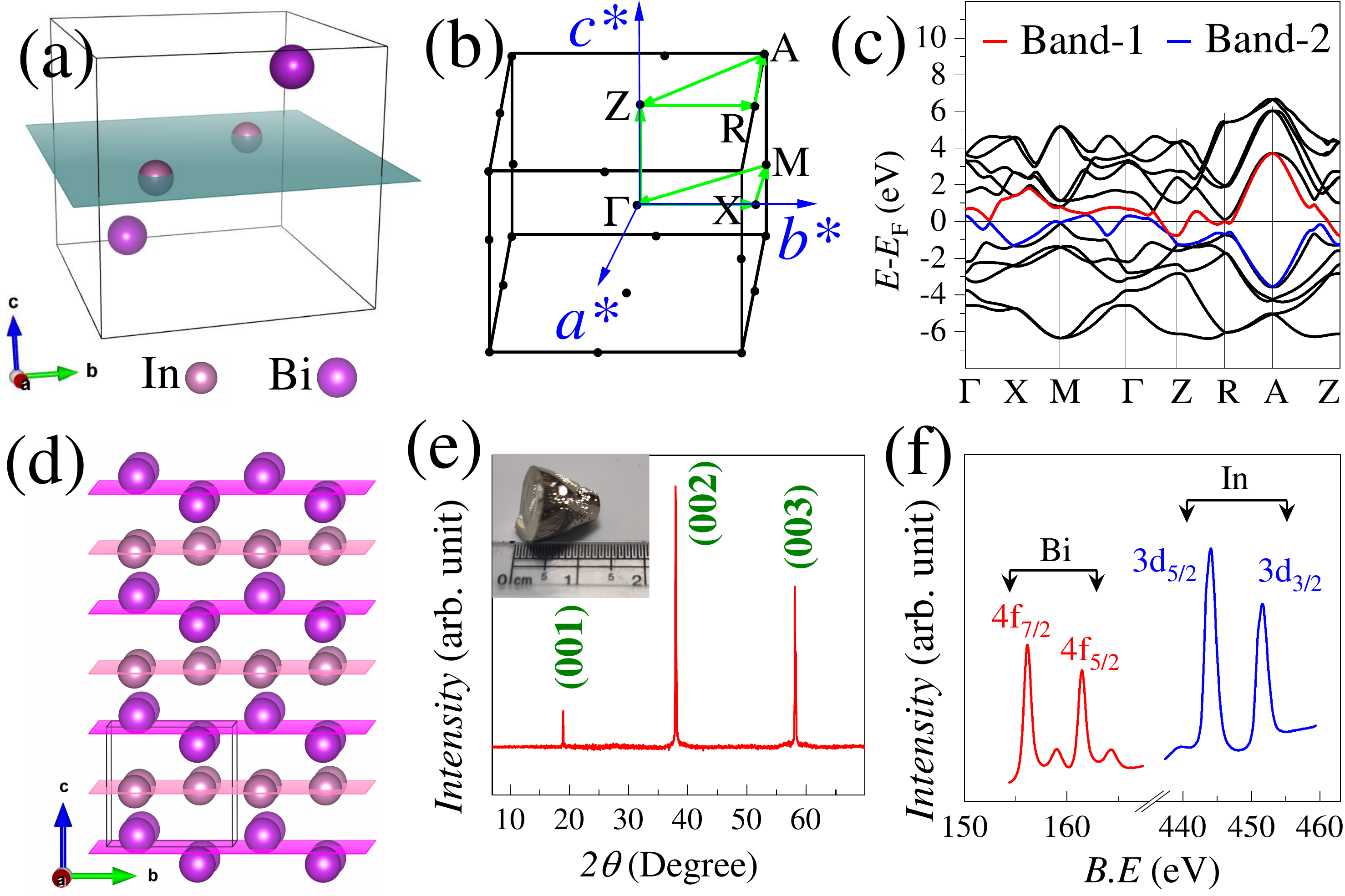}
	\caption{(a) The unit cell of InBi compound. (b) The BZ structure and the high symmetric points along which DFT calculation is performed. (c) The bandstructure along high symmeyric points. The band-1 (red) and band-2 (blue) are only two bands that contribute at $E_F$. (d) The \textit{ab}-planes of InBi. (e) XRD that performed on \textit{c}-axis. The XRD peaks are coming from the \textit{ab} plane. (Inset) The photograph of conical shape InBi single crystal. (f) The XPS spectra of InBi}
	\label{structure}
\end{figure}
We performed a powder XRD to obtain the crystal structure of the compound. Our study suggests that the compound crystallizes in \textit{p4/nmm} (Space group no. 129) crystal structure. The refinement of the powder XRD pattern reveals \textit{a} = 5.01(2) $\AA$,\space \textit{b} = 5.012(2) $\AA$ and \textit{c} = 4.77(9) $\AA$. The unit cell of the crystal indicating all atoms is shown in FIG. \ref{structure} (a). The shaded area in the unit cell indicates a plane where all In atoms lie. The  Brillouin zone (BZ) structure indicating high symmetric points and corresponding electronic  bandstructure is shown in FIG. \ref{structure} (b) and (c). The bandstructure indicates that only two bands crossed the Fermi level. We assigned these two bands as Band-1 and band-2. Band-1 and band-2 are highlighted in red and blue colour in FIG. \ref{structure} (c). The layered crystal structure of the compound is shown in FIG. \ref{structure} (d). The shaded planes indicate the \textit{ab}-plane that contain In and Bi atoms. FIG. \ref{structure} (e) shows the XRD pattern taking from the \textit{ab}-plane of the crystal. As the diffraction occur only among the \textit{ab}-planes, the diffractogram only contains $(00l)$ peaks. The very sharp peaks of the XRD pattern indicate the good crystallinity of the compound. The $(00l)$ peaks are also consistent with the space group and unit cell parameters derived from the powder XRD. We shown the photograph of our conical-shaped single crystal along with a centimeter scale in the inset of FIG. \ref{structure} (e). The X-ray photoelectron spectroscopy (XPS) of InBi is shown FIG. \ref{structure} (f). The spectra from $4f$ and $3d$ orbitals of Bi and In atoms well verified the compound's elemental composition. The XPS spectra also agree with the experimental stoichiometric ratio of the elemental composition.

\subsection{Magnetotransport properties}
\label{B}
\subsubsection{Magnetic field dependent $\rho (T)$}

\begin{figure*}
\includegraphics[width=0.98\textwidth]{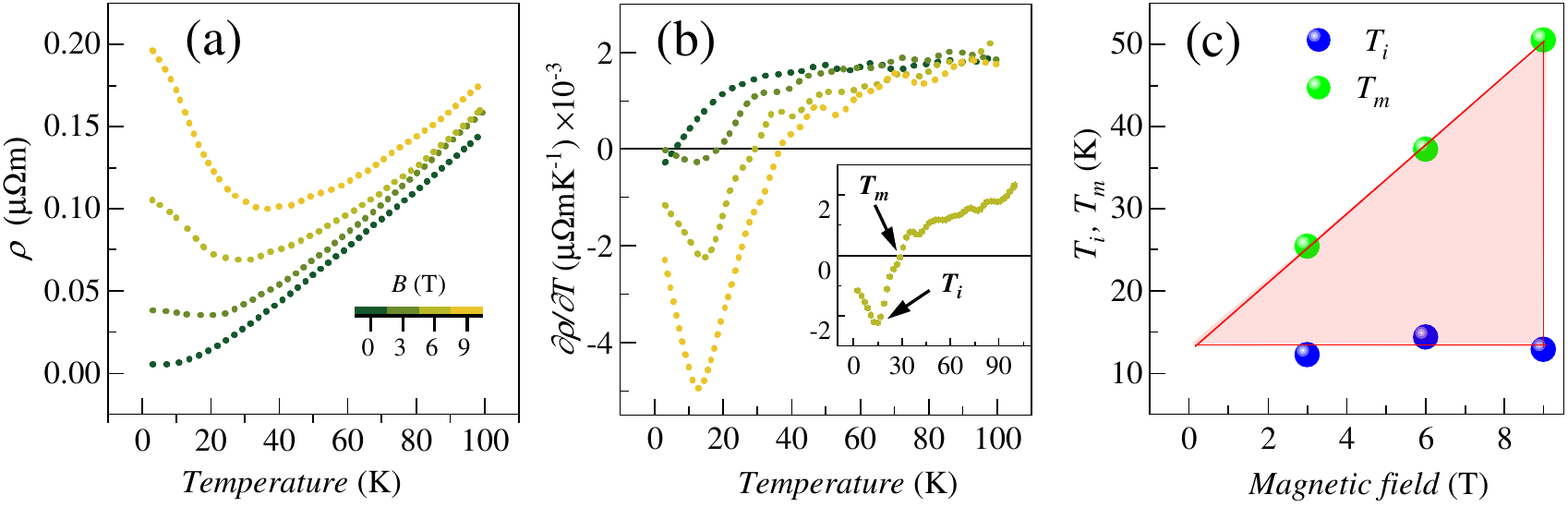}
\caption{(a) $\rho (T)$ measured at 3--100 K with the application of 0--9 T magnetic field. (b)After performing $\frac{\partial\rho (T)}{\partial T}$ (to calculate turn-on temperature). The $T_m$ and $T_i$ are indicated in the inset figure. (c) The variation of $T_m$ and $T_i$ with the magnetic field. Triangle formed by $T_m$ and $T_i$ are shown by a pink shaded area.}
\label{fig:RT1}
\end{figure*}
\begin{figure*}
	\includegraphics[width=0.98\textwidth]{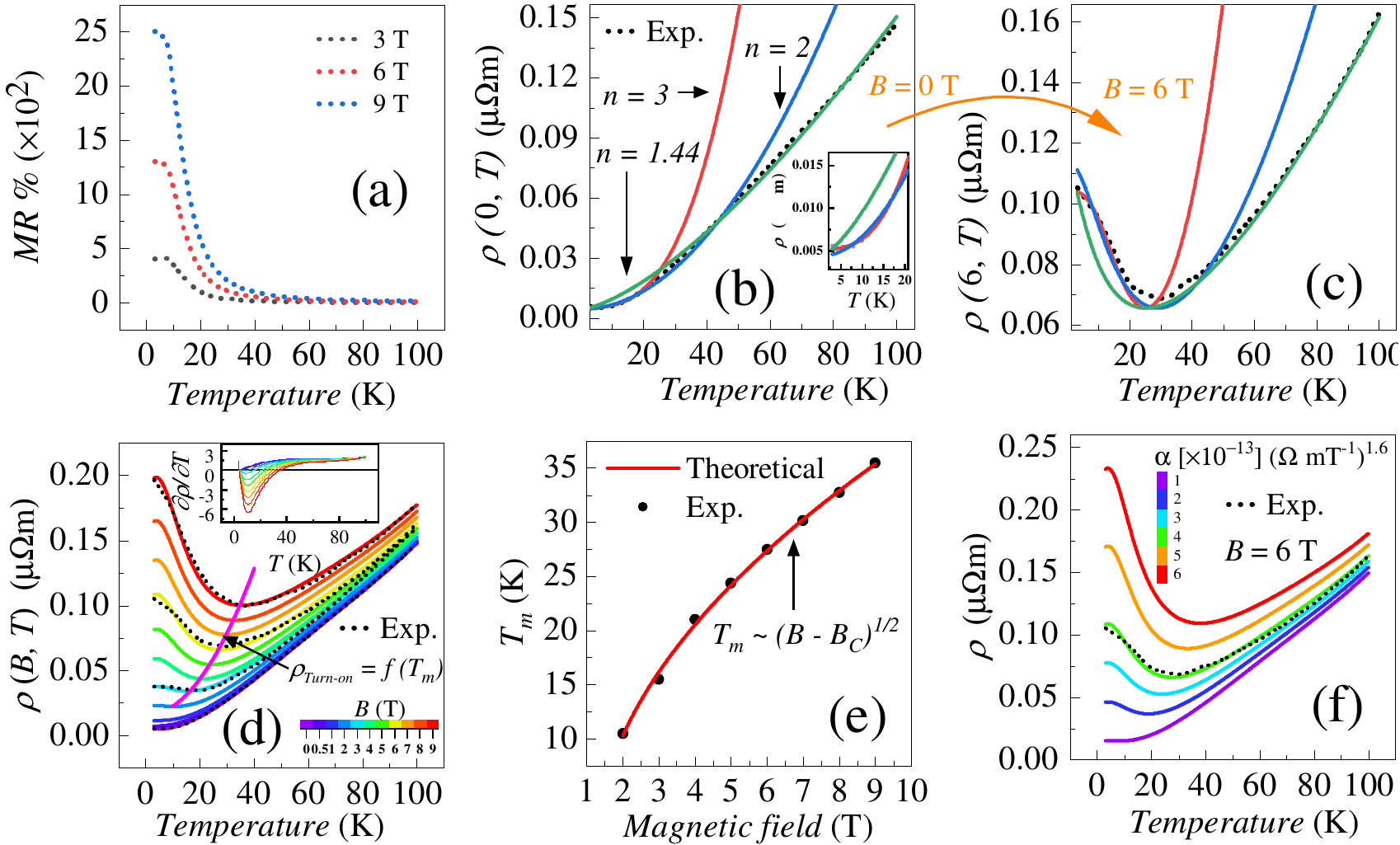}
	\caption{(a) Calculated $MR\%$ with the application of 3 T, 6 T and 9 T magnetic field. (b) Experimental $\rho (T)$ without any magnetic field and differnt type of fitted functions. Data upto 20 K is shown in the inset figure. (c) Experimental and simulated  $\rho (T)$ with the application of the 6 T magnetic field. (d) Simulated $\rho (T)$ under the application of 0--9 T magnetic field. Experimental  $\rho (T)$ (shown as black dashed lines) are plotted over the theoretically generated one. The $\rho _{turn-on}$ as a function of $T_m$ is indicated by a solid magenta line. The $\frac{\partial\rho (T)}{\partial T}$ of the theoretically generated patterns are shown in the inset figure. (e) Variation of turn-on temperature as a function of \textit{B} and the fitted pattern. (f) Simulated $\rho (6, T)$ with the variation of $\alpha$}.
	\label{fig:RT2}
\end{figure*}

 \begin{figure}
	\includegraphics[width=0.49\textwidth]{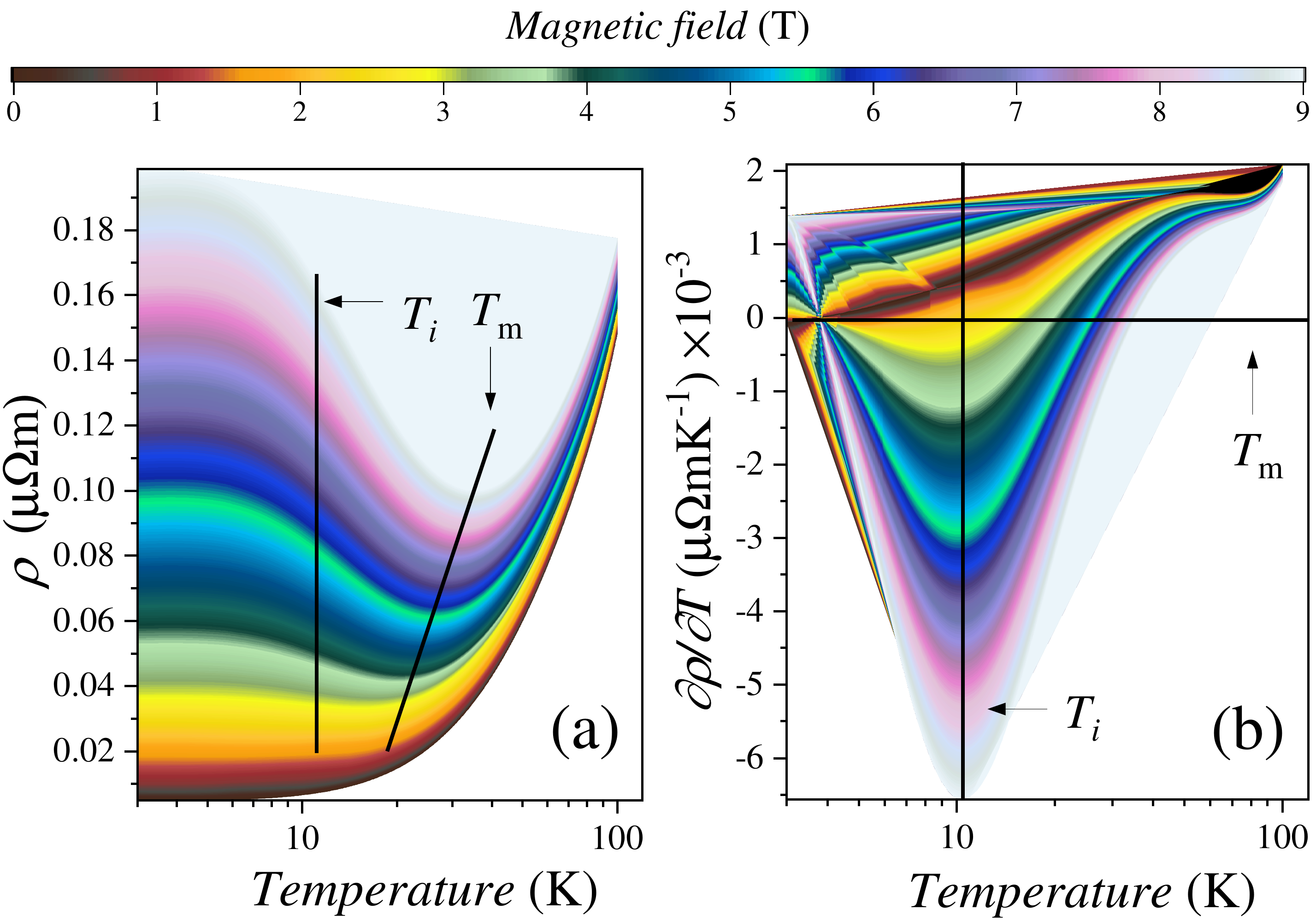}
	\caption{3D contour plot of simulated data. (a) $\rho(B,T)$ (b) $ \frac{\partial[\rho (B,T)]}{\partial T} $   }
	\label{3d_contour}
\end{figure}

The temperature dependent resistivity [$\rho(T)$] measured at 3--100 K is shown in FIG.\ref{fig:RT1}(a). The resistivity of the compound at 3 K and 100 K are 0.5$\times$10$^{-2}$ and 14.7
$\times$10$^{-2}$ $\mu$$\Omega$m, respectively, which is very close to the earlier reported result\cite{okawa2018extremely}.
To calculate the RRR of our compound, we performed our resistivity study upto 300 K. The RRR of the compound based on the 3 and 300 K resistivity data is estimated to be 93. The RRR of the compound in the earlier reported data lies between 91-271\cite{okawa2018extremely}. Our study suggests that the RRR of our compound lies within the range of earlier reported data. From FIG.\ref{fig:RT1}(a), we have observed that the resistivity of the compound sharply increases with the increase of magnetic field, especially in the  low-temperature regime. Whene \textbf{B} = 0, resistivity of the compound have been increased monotonically with temperature. On applying the high magnetic field, the $\rho(T)$ rapidly decreases first, and then it increases with the increase of temperature. Such kind of up-turn behaviour is also observed in other non-trivial topological systems\cite{shekhar2015extremely, hu2016pi, tafti2016resistivity, sun2016new, wang2015origin}, where applying certain magnetic field the  $\rho(T)$ shows minima at a particular temperature ($T _{m}$).

We have shown the 1st order derivative $ \frac{\partial[\rho (T)]}{\partial T} $ in FIG.\ref{fig:RT1}(b). Turn-on temperature is indicated in the figure as T$ _{m}$, where the slope of the $\rho(T)$ changes its sign from negative to positive. Apart from turn-on behaviour, we have also identified the minima of the $ \frac{\partial[\rho (T)]}{\partial T} $, indicated as  $T _{i}$ in FIG.\ref{fig:RT1}(b). The  $T _{i}$ of the compound is approximately estimated as 12 K for our compound, and it changes negligibly with the change of applied magnetic field. We have plotted the variation of $T _{m}$ and  $T _{i}$ with the applied magnetic field in FIG.\ref{fig:RT1}(c). The triangle formed by the  $T_{m}$ and  $T_{i}$ is one of the vital features observed in the compensated semimetals\cite{tafti2016temperature}. The value of the slope inside the shaded area of the triangle is negative, whereas the upper outside region of the triangle shows the positive slope. For the case of graphite and bismuth, the triangular phase diagram can be explained by the inequality $\hslash /\tau \lesssim \hslash\omega _c \lesssim k_{\rm B}T$, where $\tau$ is the electron-phonon scattering time and $\omega _c$ is the cyclotron frequency\cite{du2005metal}. In clean semimetals with low carrier density $\tau ^{-1} \ll k_BT/\hslash$, and hence, there exists a wide temperature-field range where XMR and turn-on like behaviour in resistivity appear. On the other hand, large carrier density and strong impurity scattering limit the MR in conventional metals.
 
We have calculate the T-dependent MR from the $\rho(T)$ data. The MR of the compound defined as 
 \begin{equation}
 	\label{eq_MR}
 	MR = [\rho(B,T) - \rho(0,T)]/\rho(0,T)
 \end{equation}
Here, $\rho(B,T)$ denoted the resistivity of the compound as a function of magnetic field (\textit{B}) and temperature (\textit{T}). The compound's $MR\%$ at 3-100 K is shown in FIG.\ref{fig:RT2}(a). The $MR\%$ of the compound is very high at low temperatures. It goes up to 2500\% at 9 T. We have observed that  $MR\%$ of the compound reduces drastically from 0 to 20 K and it almost becomes negligible above 40 K.

  To observe the temperature dependency of  $\rho(0,T)$, we have fitted $\rho(0,T)$ with several functions [FIG.\ref{fig:RT2}(b)]. First, to observe the temperature dependency at low temperature, we have fitted $\rho(0,T)$ as $A + BT^{n}$ (where $A$ and $B$ are the constant) up to 20 K. The result from the best-fitted parameter gives $n$ = 3. In the second step, we fixed $n$ = 2, and fitted the  $\rho(0,T)$ up to 20 K. However, the fitted data using $n$ = 2 is not well-matched compared to the previous case. Lastly, we have fitted $\rho(0,T)$ in the whole temperature range, and the best-fitted parameter for that scenario gives $n$ = 1.44. In the last case, the fitted data is well-matched in high temperature, but it does not fit well in low-temperature region.
  The observation of $n$ = 3 for our case indicates the departure of pure electron dominated scattering. Generally, $n$ = 2 is observed for pure electron-correlated dominated scattering\cite{ziman2001electrons}.
  Similar kinds of behaviours (n = 3) are also observed for other compensated semimetals \textit{viz}., LaSbTe\cite{singha2017magnetotransport} and LaBi\cite{singha2017fermi}. Semimetal LaSb ($n$ = 4)\cite{tafti2016resistivity}, elemental yttrium\cite{hall1959electrical} and transition-metal carbide\cite{zhang2014s} also shown interband electron-phonon scattering.     
  
  \subsubsection {origin of turn-on behaviour}
  
  The MR of the compound based on the Kohler's law\cite{kohler1938magnetischen, xu2021extended} is defined as
  \begin{equation}
  	\label{eq_kohler}
  	MR = \alpha[B/\rho(0,T)]^m
  \end{equation} 

Putting the Eq. \ref{eq_MR} in to the Eq. \ref{eq_kohler}, we get

\begin{equation}
	\rho (B,T) = (\alpha B^{m}/[\rho(0,T)]^{m-1}) + \rho(0,T)
	\label{eq_RT}
\end{equation}

Using Eq. \ref{eq_RT}, we have simulated the $\rho(T)$ under the 6 T magnetic field. During the simulation, we have fixed $\alpha$ = 1.2$\times$10$^{-13}$ ($\Omega$mT$^{-1}$)$^{1.6}$ and \textit{m} = 1.6. Ideally, \textit{m} = 2 for the perfectly compenseted system but, our result \textit{m} = 1.6 suggests that our compound slightly departs from the perfectly compnsated case. The simulated patterns of $\rho(6,T)$ along with the experimental data are shown in FIG. \ref{fig:RT2} (c). Our study suggests that simulated $\rho(6,T)$, derived from the experimental $\rho(0,T)$ reasonably matches the experimental $\rho(6,T)$. we have also predicted $\rho(6,T)$ for all the corresponding cases of $\rho(0,T)$. Our simulated $\rho(6,T)$ in FIG. \ref{fig:RT2} (c) follows the same colour code for the corresponding cases of  $\rho(0,T)$ in FIG. \ref{fig:RT2} (b). The result suggests that for $n$ = 3 and 2, the theoretically generated $\rho(6,T)$ matches well at low temperature but deviates at high temperature. For the case of $n$ = 1.44, simulated $\rho(6,T)$ fairly matches with experimental one at high temperature.  

Using Eq. \ref{eq_RT}, we have simulated  $\rho(B,T)$ for 0.5--9 T. We have plotted the theoretically generated  $\rho(B,T)$ in FIG. \ref{fig:RT2} (d). The result matches the experimental data excellently. The experimental data are plotted as a black dashed line over the corresponding simulated pattern. Such an analysis strongly validates our model that finely replicate our unique  $\rho(B,T)$. We also shown the first order derivative of $\rho(B,T)$ in the inset of FIG. \ref{fig:RT2} (d) to calculate the turn-on temperature ($T_m$). Derived $T_m$ with the corresponding magnetic field are potted in  FIG. \ref{fig:RT2} (e). The variation of turn-on temperature with respect to the magnetic field is fitted with the equation
\begin{equation}
	T_m = \zeta (B - B_c)^\nu
	\label{eq_1}
\end{equation}
Such kind of formulation is theoretically predicted by D. V. Khveshchenko\cite{khveshchenko2001magnetic}, and validated for graphene and bismuth. For the graphite and bismuth the $\nu$ = 1/2. The best fitted parameters in our case give $\zeta$ = 12.75 K.T$^{-0.5}$, $B_c$ = 1.36 T and $\nu$ = 1/2. $\nu$ = 1/2 indicates that the turn-on nature of our compound follows Khveshchenko's rule.

In FIG. \ref{fig:RT2} (f), we simulate $\rho (B, T)$ with the variation of $\alpha$. In the simulation, we fixed \textit{B} = 6 T, and observe how the variation of $\alpha$ affects the turn-on behaviour. Our analysis suggests that the lower the $\alpha$, the higher the critical magnetic field needed to produce turn-on behaviour. Our analysis indicates that if $\alpha$ $ < $ 0.48$\times$10$^{-13}$ ($\Omega$mT$^{-1}$)$^{1.6}$, we could not observe any turn-on behaviour even at 6 T. The analysis suggests that there is an enormous role of $\alpha$ in producing turn-on phenomena.

For further studies , we give some mathematical analysis regarding to the turn-on behaviour. First, We define the resistivity minima at $T_m$ as $\rho _{turn-on}$. To calculate the minima in $\rho(B,T)$, we take $ \frac{\partial[\rho (T,B)]}{\partial T} $ and make it zero at $T = T_m$. The mathematics is followed as  

\begin{equation}
	\label{eq_derivetive}
\frac{\partial[\rho (B, T)]}{\partial T} \arrowvert _{T = T_m} = 0
\end{equation}

 Solving the Eq. \ref{eq_derivetive}, we get
 
 \begin{equation}
 \rho (0, T_m) = \alpha^{1/m} (m-1)^{1/m}B
 \label{eq_final_1}
 \end{equation}
To calculate the resistivity at turn-on temperature, we put $T = T_m$ in Eq. \ref{eq_RT}. The equation follows as:

\begin{equation}
	\rho (B, T_m) = \alpha^{1/m} (m-1)^{1/m}\left( \frac{1}{m-1}+1\right)B
	\label{T_m} 
\end{equation}

From Eq. \ref{eq_1}, we already know that the parameter $T_m$ is a explicit function of \textit{B}. Hence, putting the value of \textit{B} from Eq. \ref{eq_1} to Eq. \ref{T_m}, we get $\rho _{turn-on}$ as a function of $T_m$. After rewriting Eq. \ref{T_m}, we get
\begin{equation}
	\rho _{turn-on} (T_m) = \alpha^{1/m} (m-1)^{1/m}\left(\frac{1}{m-1}+1\right)\left(\frac{T_m^2}{\zeta^2} + B_c\right)
	\label{B_c1}
\end{equation}

Eq. \ref{B_c1} is the general formula we developed which valid for particular compound that shows turn-on nature in $ \rho (T)$ on the application of magnetic field. Putting the value of $\alpha$ = 1.2$\times$10$^{-13}$ ($\Omega$mT$^{-1}$)$^{1.6}$, $m$ = 1.6, $\zeta$ = 12.75 K.T$^{-0.5}$ and B$_c$ = 1.36 T, Eq. \ref{B_c1} in our case becomes

\begin{equation}
	\rho _{turn-on} (T_m) = \left[\frac{T^2_m}{163} + 1.36\right] \times 10^{-8}
\label{B_c3}
	\end{equation}
 
 We plot the Eq. \ref{B_c3} in Fig. \ref{fig:RT2} (d). The magenta line in the figure indicates the $\rho _{turn-on}$. The simulated $\rho _{turn-on}$ based on Eq. \ref{B_c3} excellently matches with the experimental results. \\

 From the whole transport study, we want to give some conclusive remarks. First, without the application of the magnetic field resistivity of the compound follows as A +B$T^n$. At low temperature \textit{n} = 3. But at higher temperature\textit{ n} becomes 1.44. Secondly, based on Eq. \ref{eq_RT}, we have generated a complete two dimensional data set (contains \textit{B} and \textit{T} ) that fully capable of replicating the experimental data. We shown the simulated resistivity data and its 1st derivative in a 3D contour plot in FIG. \ref{3d_contour} (a)-(b). Thirdly, the parameter $\alpha$ has a great role in producing the turn-on behaviour in $\rho (T)$. Turn-on phenomena can be completely vanished if $\alpha$ becomes lower than critical value. Finally, to calculate the resistivity at turn-on temperature ($T_m$), we developed a generalized formulation that finely simulate the experimental result. The mathematical formulation based on Eq. \ref{B_c1} is solely developed by ourselves.         
   
    \subsubsection {Temperature dependent $\rho (B)$}
    
    \begin{figure*}
    	\includegraphics[width=0.98\textwidth]{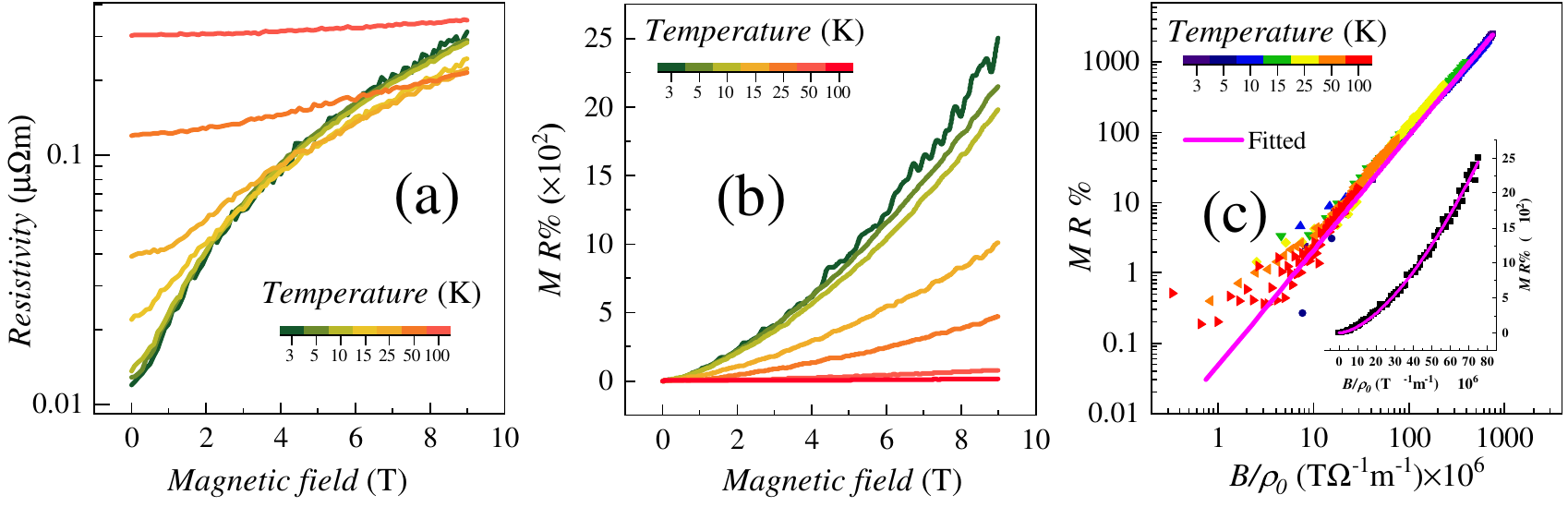}
    	\caption{(a) The $\rho (B)$ measured at 3--100 K with the application of 0--9 T magnetic field. (b) The $MR\%$ as a function of \textit{B} that is calculated from the experimental  $\rho (B)$. (c) The $\log-\log$ plot of $MR\%$ as a function of $B/\rho (0, T)$. The fitted line based on the Kohler's law is shown in a solid magenta line. The fitted line for 3 K is separately shown in the inset figure.}
    	\label{kohler}
    \end{figure*}
    
    The variation of $\rho (B)$ at several temperatures is shown in FIG. \ref{kohler} (a). We have shown that the change of resistivity with temperature is very high when applied magnetic field is very low and $\rho (B)$ of all temperatures trend to converge in a singular region at higher magnetic field. It is observed that with the increase of \textit{B}, $\rho (B, 3)$ is growing firstly, and it almost catches up the value of $\rho (B, 15)$, $\rho (B, 25)$ an $\rho (B, 50)$ at 3 T, 4 T and 6 T respectively. The\textit{ MR\%} as a function of \textit{B} is shown in FIG. \ref{kohler} (b). The \textit{ MR\%} of the compound is very high below 10 K and reaches up to 2500\% at 3 K. On the other hand, \textit{MR\%} at higher temperature reduces drastically and almost becomes negligible above 25 K. We have also shown \textit{MR\%} \textit{vs} $B/\rho(T, 0)$ for all the temperatures in FIG. \ref{kohler} (c). The $\log-\log$ plot of all the data points falls in a single straight line. The fitted line based on the Eq. \ref{eq_kohler} is shown by a solid magenta line. During the fitting, we have put corresponding value of $\alpha$ and $ m $ as stated earlier. The fitted data at 3 K before the $\log-\log$ plotting is shown in the inset of FIG. \ref{kohler} (c).   
    The falling of all data points in a singular straight line also verified our earlier assumption that the parameter $\alpha$ and $ m $ are almost temperature-insensitive for our compound.
    
      \subsubsection{Hall resistivity study}
      
      \begin{figure}
      	\includegraphics[width=0.49\textwidth]{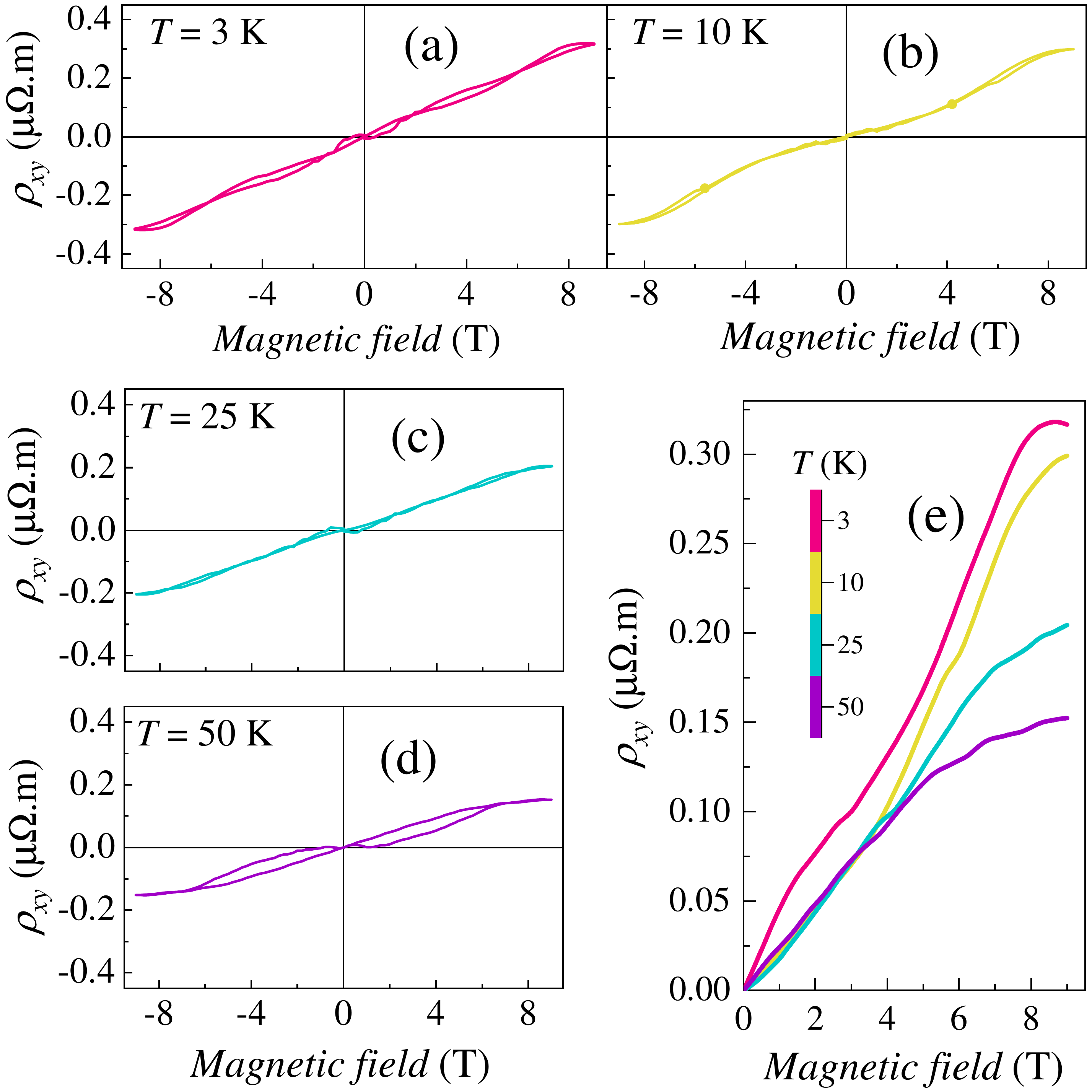}
      	\caption{The $\rho_{xy} (B)$ from +9 T to -9 T (four-quadrant plot) at (a) 3 K, (b) 10 K, (c) 25 K and (d) 50 K. (e) The $\rho_{xy} (B)$ from 0--9 T with the temperature range 3--50 K}
      	\label{fig:Hall}
      \end{figure} 
  \begin{figure}
      	\includegraphics[width=0.49\textwidth]{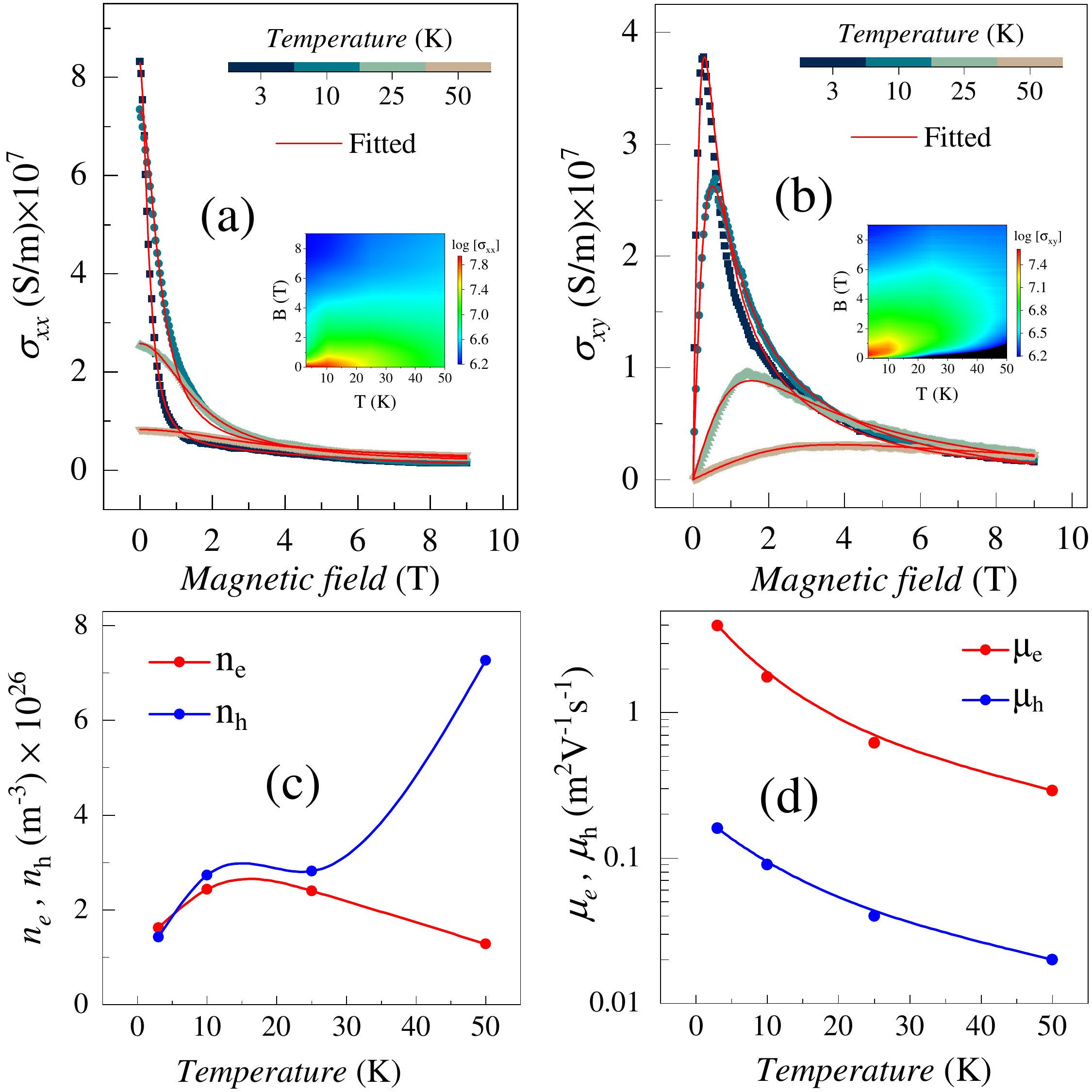}
      	\caption{(a) $\sigma _{xx} (B)$ and its corresponding fittings, (inset) logarithmic plot of $\sigma _{xx} (B)$ as a function of magnetic field and temperature in a 3D contour plot. (b) $\sigma _{xy} (B)$ and its corresponding fittings, (inset) logarithmic plot of $\sigma _{xy} (B)$ as a function of magnetic field and temperature in a 3D contour plot. (c) The temperature-dependent electron and hole density that is estimated from the two carrier model. (d) The temperature-dependent electron and hole mobility that is estimated from the two carrier model.  }
      	\label{fig:sigma}
      \end{figure}
     
 The Hall effect study in 0--9 T magnetic field at different temperature are shown in FIG. \ref{fig:Hall}. From the figure, $\rho _{xy} (B)$ shows a nonlinear nature at all temperatures. Such kind of behaviour suggests that both carriers (electron and hole) are present in the transport phenomena. We have also observed that at T $<$ 25 K, $\rho _{xy} (B)$ increases rapidly. It is shown that on application of 9 T magnetic field, $\rho _{xy}$ at 3 K is almost twice comparing to the  $\rho _{xy}$ at 25 K. Interestingly, the change in \textit{ MR\%} calculated from  $\rho _{xx}$ also twice in the same temperature and magnetic field.
 
 Using experimental result of $\rho _{xx} (B)$ and $\rho _{xy} (B)$, we have calculated the $\sigma _{xx} (B)$ and  $\sigma _{xy} (B)$. After the inverting of the resistivity matrix, $\sigma _{xx} (B)$ and  $\sigma _{xy} (B)$ can be written as
 \begin{equation}
 	\sigma _{xx} (B) = \frac{\rho _{xx} (B)}{[\rho _{xx} (B)]^2 + [\rho _{xy} (B)]^2}
 	\label{eq_sigma_xx}
\end{equation} 

 \begin{equation}
	\sigma _{xy} (B) = \frac{\rho _{xy} (B)}{[\rho _{xx} (B)]^2 + [\rho _{xy} (B)]^2}
		\label{eq_sigma_xy}
\end{equation} 

Based on the Eq. \ref{eq_sigma_xx} and \ref{eq_sigma_xy}, calculated $\sigma _{xx} (B)$ and  $\sigma _{xy} (B)$ are plotted in FIG. \ref{fig:sigma} (a) and \ref{fig:sigma} (b). 

We have fitted $\sigma _{xx} (B)$ and  $\sigma _{xy} (B)$ with two band two carrier model\cite{hurd2012hall} stated as
\begin{equation}
\sigma _{xx} (B) = e\left[n_h\mu_h \frac{1}{1 + (\mu_h B)^2} + n_e\mu_e \frac{1}{1 + (\mu_e B)^2}\right]
\label{eq1}
\end{equation}

\begin{equation}
	\sigma _{xy} (B) = eB\left[n_h\mu_h^2 \frac{1}{1 + (\mu_h B)^2} - n_e\mu_e^2 \frac{1}{1 + (\mu_e B)^2}\right]
    \label{eq2}
\end{equation}

 As we stated earlier that both the carrier types contribute to our transport phenomena, we have performed the global fitting based on the two carrier model where simultaneous presence of electron and hole carrier is considered. Here electron density ($n_e$), hole density  ($n_h$), electron mobility ($\mu _e$) and hole mobility ($\mu _h$) are termed as a fitting parameters. The fitted lines are shown as a solid red line in FIG. \ref{fig:sigma} (a) and \ref{fig:sigma} (b). Extracted parameters from the Eq. \ref{eq1} and \ref{eq2} give almost same result. The fitting parameters \textit{viz}., $n_e(T)$, $n_h(T)$, $\mu _e(T)$ and $\mu _h(T)$ are shown in FIG. \ref{fig:sigma} (c) and \ref{fig:sigma} (d).
 
 From FIG. \ref{fig:sigma} (c), it is observed that $n_e(T)$ is slightly higher at 3 K but, as a broad view the $n_e(T)$ and $n_h(T)$ are approximately same ($\sim$ 1.5$\times$10$^{-26}$ m$^{-3}$) below 25 K. After 25 K, $n_e(T)$ almost remains the same and $n_h(T)$ increases to $\sim$ 7$\times$10$^{-26}$ m$^{-3}$. The results suggest that the number of electrons and holes contributes equally at lower temperature whereas, at higher temperatures, the transport phenomena are slightly dominated by holes. Observing the MR data from FIG. \ref{kohler} (b), we observed that MR\% also very high bellow 25 K. Such kind of correlation indicates that the phenomena XMR is accompanied by the electron-hole compensation of the compound. Regarding to the carrier mobility, we have observed that the electron mobility of the compound at 3 and 15 K are 3.98 and 0.29 m$^2$V$^{-1}$s$^{-1}$ respectively. The hole mobility at 3 and 15 K are 25 and 15 times lower than the corresponding electron mobility at that temperature. The effective mobility ($\mu _{eff}$ $\sim$$\sqrt{\mu _e  \mu _h}$) at 3 K is 0.79 m$^2$V$^{-1}$s$^{-1}$, which is consistent with the earlier reported data\cite{okawa2018extremely}.

 \subsection{Shubnikov-de Haas oscillation and Fermi surface topology study}
 \label{C} 
 \begin{figure*}
 	\includegraphics[width=0.98\textwidth]{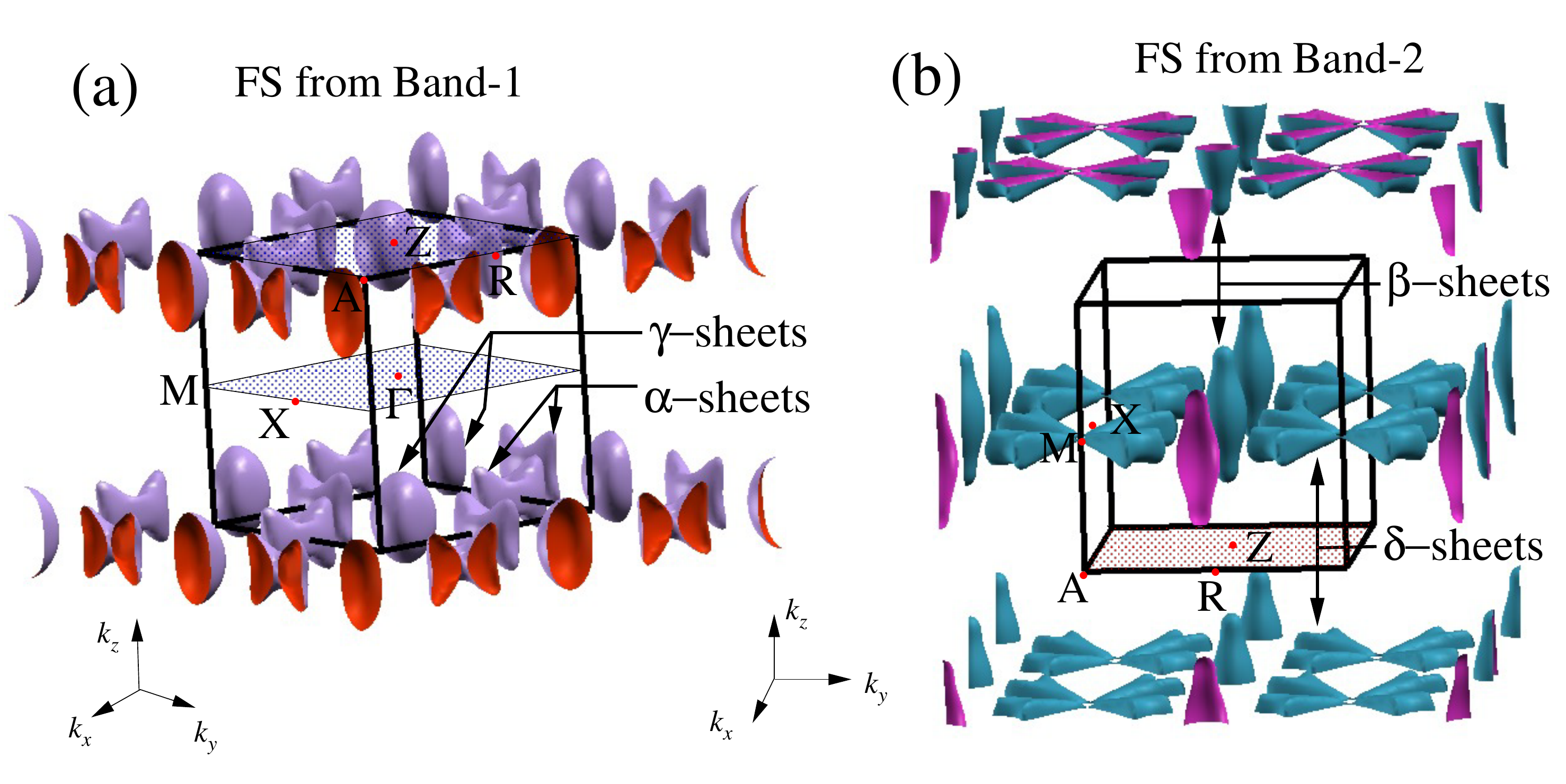}
 	\caption{(a) The electron pockets come from band-1. The $\gamma$ and $\alpha$ sheets are lying at high symmetric Z and R points respectively. (b) The hole pockets come from band-2. The $\beta$ and $\delta$ sheets are lying at high symmetric $\Gamma$ and M points respectively.}
 	\label{fs_extended}
 \end{figure*}
 
 \begin{figure*}
 	\includegraphics[width=0.98\textwidth]{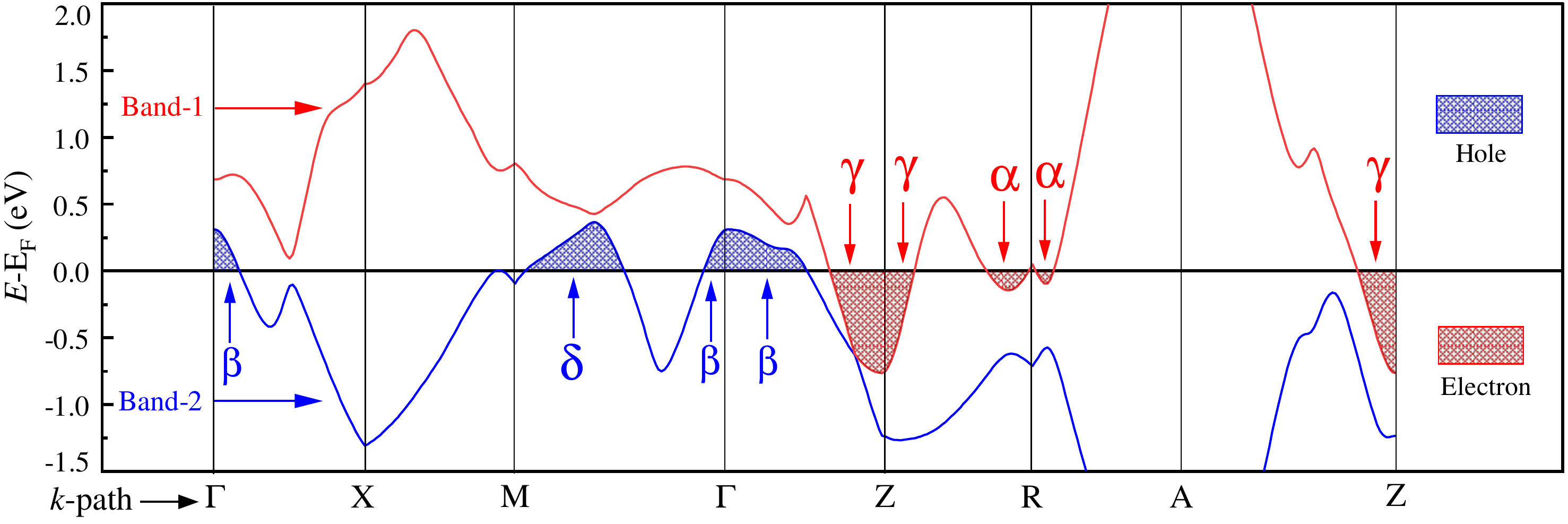}
 	\caption{The bandstructure of the band-1 is shown in red. The allowed states marked by red shaded area denoted the electron-like pocket. The $\gamma$ and $\alpha$ pockets are generated from the band-1. The bandstructure of the band-2 is shown in blue. The allowed states marked by blue shaded area denoted the hole-like pockets. The $\beta$ and $\delta$ pockets are generated from the band-1.}
 	\label{bandstructure}
 \end{figure*}
 
 \begin{figure*}
 	\includegraphics[width=0.98\textwidth]{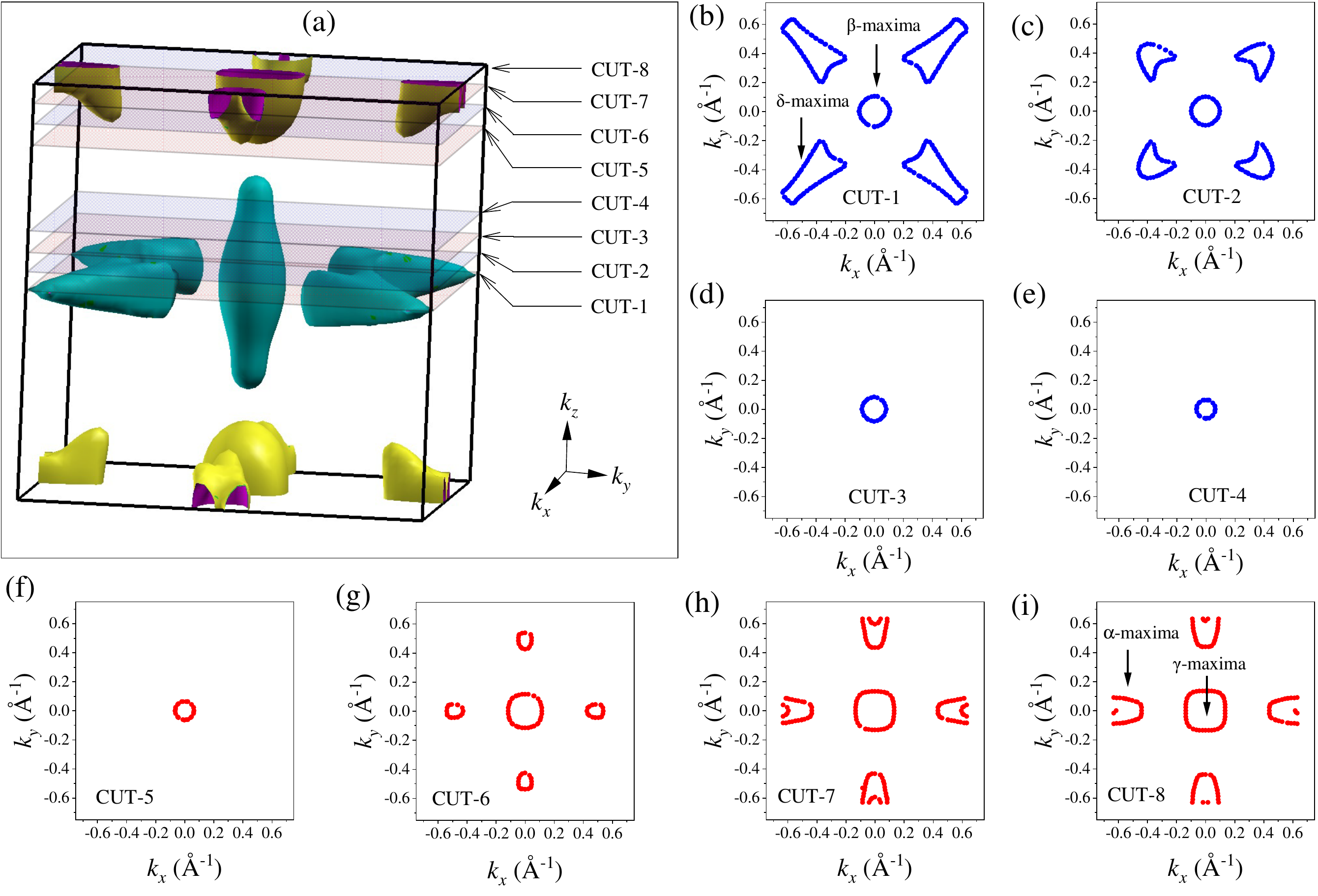}
 	\caption{Fermi sheets in \textit{k$_x$-k$_y$} plane at several \textit{k$_z$} cut. (a) Several \textit{k$_x$-k$_y$} plane in 3D BZ where the 2D contours are plotted. (b)-(d) The contours in blue coming from band-2. the areas of $\beta$ and $\delta$ sheets are continuously decreasing from CUT-1 to CUT-4. The arrow mark in CUT-1 indicates the extrimum area of the corresponding pockets. (f)-(i) The contours in red coming from band-1. the areas of $\alpha$ and $\gamma$ sheets are continuously increasing from CUT-5 to CUT-8. The arrow mark in CUT-8 indicates the extrimum area of the corresponding pockets.}
 	\label{image-slice}
 \end{figure*}

 \begin{figure*}
 	\includegraphics[width=0.98\textwidth]{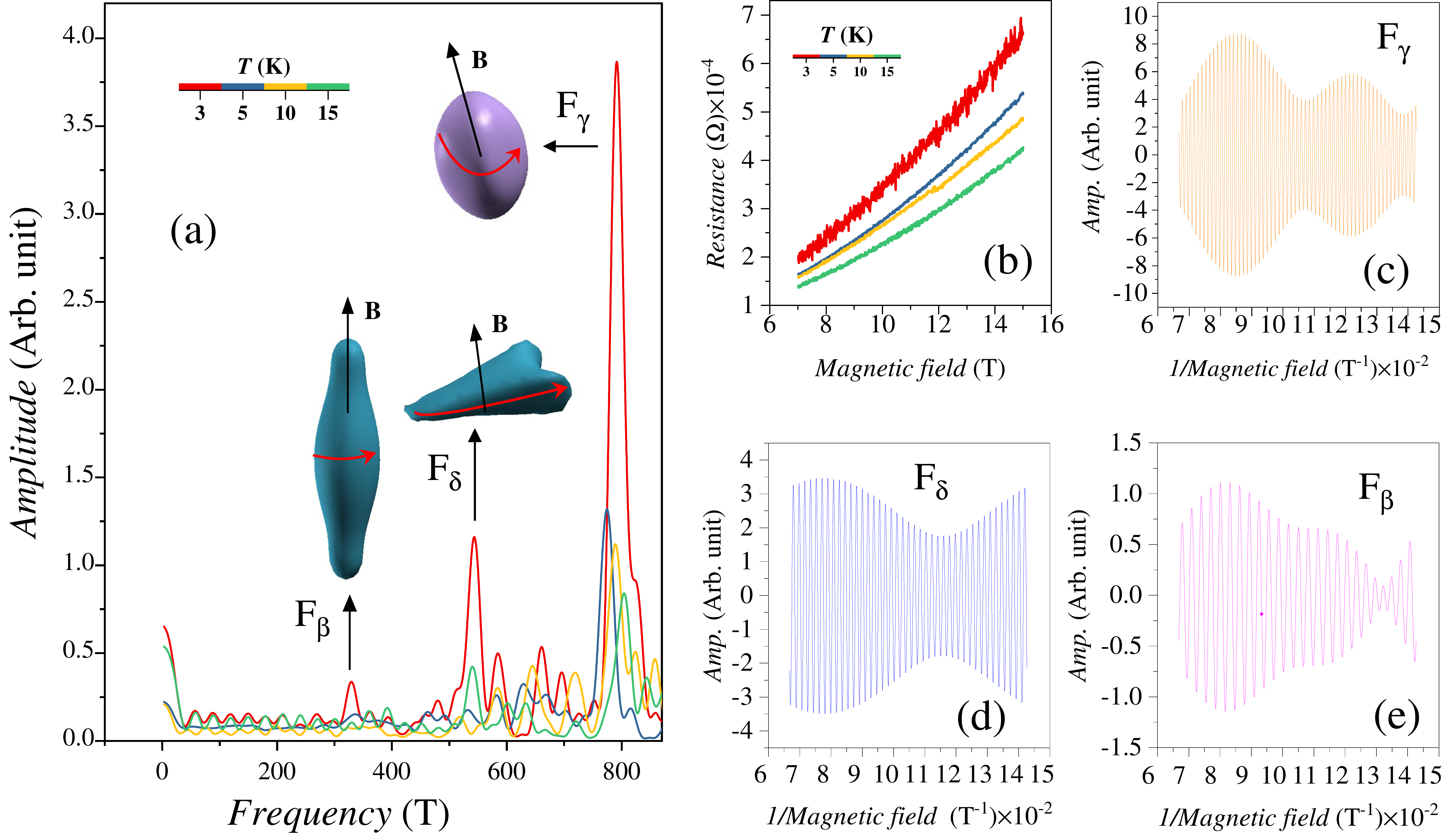}
 	\caption{The FFT spectra and SdH oscillation. (a) The FFT spectra calculated at 7-15 T field. The red arrow indicates the extrimum orbits of the corresponding pockets from where frequency peaks are coming. (b) The actual MR data from where FFT is performed. (c)-(e) The deconvoluted spectra coming from corresponding pocket at 3 K.  }
 	\label{fig:SdH}
 \end{figure*} 

\begin{figure}
	\includegraphics[width=0.49\textwidth]{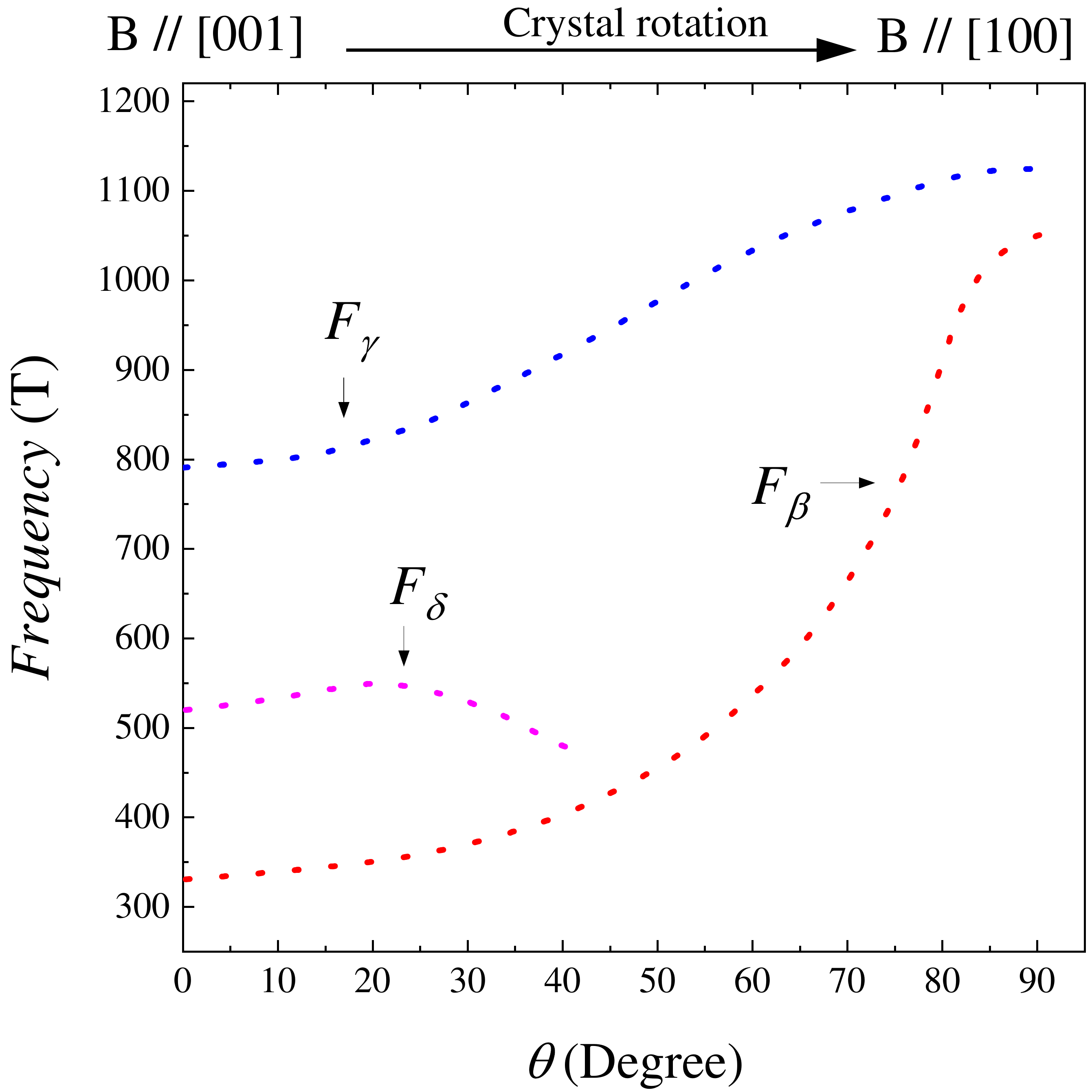}
	\caption{Data taken from Meyer \textit{et al.}\cite{meyer1974haas}. Angle dependent variation of three major frequencies $F{_\beta}$, $F{_\gamma}$ and $F{_\delta}$ }
	\label{rotation}
\end{figure}
 
 \begin{figure*}
 	\includegraphics[width=0.98\textwidth]{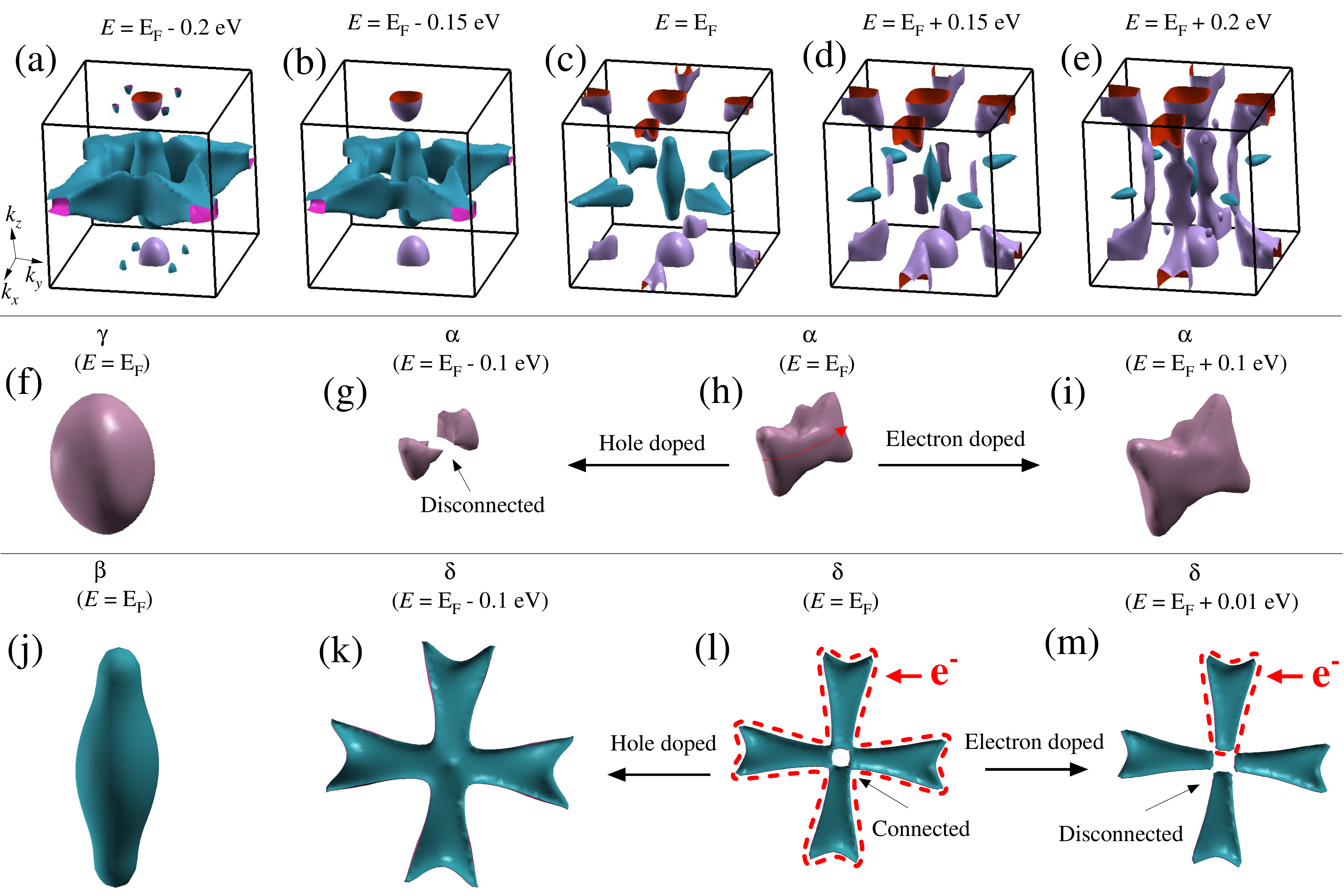}
 	\caption{The veriation of Fermi sheets from hole doping to electron doping. (a)-(e) The evolution of 3D BZ from hole doping to electron doping. (f) The shape of the $\gamma$ pocket at E = E$_F$. (g)-(i) The evolution of $\alpha$ pocket from hole doped to electron doped. The extrimum orbit parallel to \textit{k$_x$-k$_y$} plane is shown by the red arrow. The substantial hole doping split the $\alpha$ pocket by two parts. (j) The shape of the $\beta$ pocket at E = E$_F$. (k)-(m) The evolution of $\beta$ pocket from hole doped to electron doped. The extrimum orbit parallel to \textit{k$_x$-k$_y$} plane is shown by red dotted line.}
 	\label{3d-evolution}
 \end{figure*}

\begin{table*}[]
 	\caption{Fermi sheets properties}
 	\label{table-1}
 	
 	\begin{tabular*}{0.98\textwidth}{@{\extracolsep{\fill} }  c  c  c  c  c  c}

 		\toprule\toprule
 		FS pocket & Origin & Type     & \begin{tabular}[c]{@{}c@{}}Experimental SdH$^{a}$\\   Frequency (T)\end{tabular} & \begin{tabular}[c]{@{}c@{}}Experimental\\   area$^{b}$ (nm$^{-2}$)\end{tabular} & \begin{tabular}[c]{@{}c@{}}Theoretical\\ area$^{(c)}$ (nm$^{-2}$)\end{tabular} \\ \midrule
 		$\beta$   & Band-2 & Hole     & 330                                                                                              & 3.1                                                                             & 3.4                                                                            \\
 		$\delta$  & Band-2 & Hole     & 543                                                                                              & 5.1                                                                             & 5.3                                                                            \\
 		$\alpha$  & Band-1 & Electron & Merged with $\beta$                                         & --                                                                             & 3.3                                                                            \\
 		$\gamma$  & Band-1 & Eectron  & 792                                                                                              & 7.5                                                                             & 6.7                                                                            \\ \midrule
 		\multicolumn{6}{l}{\begin{tabular}[c]{@{}l@{}}(a) At 3 K and \textbf{B} along the \textit{c}-axis, (b) Area derived from our SdH frequency\\ (c) Extremum area (from our DFT calculation) in \textit{k$_x$-k$_y$} plane\end{tabular}}                                                                                                                             \\ \bottomrule\bottomrule
 	\end{tabular*}
 	\end{table*}

 \subsubsection{The FS topology of InBi}
Before introducing the experimental part, we discussed the basic structure of Fermi surface of InBi. We have shown the 3D Fermi surface in FIG. \ref{fs_extended}. Our study suggest that only two bands contribute at the Fermi level. The FS coming out from band-1 creates two distinct pockets. In FIG. \ref{fs_extended} (a), we marked these pockets as $\gamma$  and $\alpha$. The $\gamma$ and $\alpha$ pockets are placed at high symmetric points Z and R respectively. Similarly, the FS coming out from band-2 also construct two pockets. We have termed these two pockets $\beta$ and $\delta$ as indicated in FIG. \ref{fs_extended} (b). The $\beta$ and $\delta$ pockets lie at high symmetric points $\Gamma$  and M respectively. Energy dispersion of the band-1 and band-2 are shown in FIG. \ref{bandstructure} where allowed electron and hole-like states are shown in red and blue shaded areas respectively. FIG. \ref{bandstructure} indicates that $\gamma$ and $\alpha$-pockets coming from band-1 are electron types. Similarly, $\beta$ and $\delta$-pockets coming from band-2 are hole types.

After identifying all the electron and hole pockets, we have calculated the extremum area of these pockets. To draw the extrema, we take several cuts in $k _{x}-k _{y} $ plane in 3D BZ. The contours from several cut perpendicular to the $ k_{z} $ axis are shown in FIG. \ref{image-slice}. The red and blue contours come from band-1 and band-2 respectively. FIG. \ref{image-slice} (b)-(e) suggests that the extremum area of $\beta$ and $\delta$ sheets lie at the $\Gamma$ point. The cut corresponding to the $\Gamma$ point is shown in FIG. \ref{image-slice} (b) where maxima of the Fermi surface are indicated by a arrow. It is visible that the area of the $\beta$ and $\delta$ sheets continuously decreases from CUT-1 to CUT-4. Similarly, from CUT-5 to CUT-8, the area of the $\gamma$ and $\alpha$ sheets continuously increase. The maximum area of $\gamma$ and $\alpha$ sheets lies at Z-point (CUT-8). The extremum area of $\gamma$ and $\alpha$-sheets are indicated by the arrow in FIG. \ref{image-slice} (i). The extremum area of all four Fermi sheets is listed in TABLE-\ref{table-1}.

\subsubsection{Experimental verification}

The quantum oscillation work of InBi was performed by Y. Saito\cite{saito1962haas} in 1964. In this same work, the author performed de Haas-van Alphen (dHvA) oscillation for InBi. After Saito's work, many authors performed the quantum oscillation study, but none of them successfully resolved the nature of electron and hole pocket from the oscillatory data. In 1973 Meyer \textit{et al.}\cite{meyer1974haas} performed dHvA study of InBi in all crystallographic directions. We have shown a part of Meyer's result in FIG. \ref{rotation} as a reference. The author predicts that $F_{\beta}$ and $F_{\gamma}$  are originating from the InBi and $F_{\delta}$ is originating from the impurity band. Observing the angular variation of dHvA oscillation, Meyer \textit{et al.} first predicted that the 3D topology of the FS originated from $F_{\beta}$ and $F_{\gamma}$ should be distorted ellipsoid. From the angle-dependent data, they cannot reveal the origin of each FS from the BZ structure. We brought the Meyer's result as it plays a crucial role in the context of our current study. We started our experiment with magneto-transport studies \textit{viz.} MR at low \textit{temperature} and high \textit{magnetic field}. To get the SdH signal, we measured MR at 3 K, 5 K, 10 K, and 15 K in the \textbf{B}-field 7--15 T. The MR data and corresponding FFT spectra are shown in FIG. \ref{fig:SdH} (b) and (a) respectively. All FFT peaks named as $F_{\beta}$, $F_{\delta}$ and $F_{\gamma}$ are highlighted in FIG. \ref{fig:SdH} (a). The time scale spectra for corresponding frequencies are shown in FIG. \ref{fig:SdH} (c)-(e). We observed that there is a considerable amount of beat formation in time scale spectra. The all beats originate from the finite broadening of the FFT peaks. The formation of beats in the oscillation is an acceptable phenomena for any compound\cite{shoenberg2009magnetic}. The detail discussion of each frequency peaks is given in the subsequent section.       

\subsubsection{Frequency from $\beta$-sheets ($F_{\beta}$)} The $F_{\beta}$ of FFT spectra is observed at 330 T as shown in FIG. \ref{fig:SdH} (a).  The extremum area of the corresponding FS is close to our theoretical calculation (see the TABLE-\ref{table-1}).  As shown in FIG. \ref{3d-evolution} (j), the 3D shape of $\beta$-pocket looks like a distorted ellipsoid structure. The angle dependent study of $F_{\beta}$ by Meyer \textit{et al.} is shown by a red dashed line in FIG. \ref{rotation} It indicates that if we change the crystal orientation from [001] to [100], $F_{\beta}$ changes from 330 T to 1052 T. The angular dependency of $F_{\beta}$ suggests that the semi-major axis of the pocket is 3.3 times longer than its semi-minor axis. Such kind of experimental result is very consistent with our theoretically generated FS structure. Our theoretical calculation also suggests that if we move the Fermi level slightly upward (electron dope), the pocket size becomes lower. Our system might behave as an electron-doped compound, as it shows slightly lower FS area than the theoretical one.

\subsubsection{Frequency from $\delta$-sheets ($F_{\delta}$)} The $F_{\delta}$ of the FFT spectra is observed at 543 T as shown in FIG. \ref{fig:SdH} (a). Origin of the particular frequency was not established earlier. Our study suggests that the  $F_{\delta}$ comes from the distorted triangular shape FS lying at the M point. This FS has a four-fold symmetry as shown in FIG. \ref{3d-evolution} (k)-(m). The FIG. \ref{3d-evolution} (l) indicates that at \textit{E} = $E_{\rm F}$ the four petals are connected and electrons can move through the FS to make a complete loop. The extremum orbit of the electron path is marked by a red dashed line. There is an additional possibility that electrons can circulate within a single petal of FS as the linkage between two petals is infinitesimally small. Our study suggests that any finite upward movement of $E_{\rm F}$ can disconnect the linkage among the petals. We show an example in FIG. \ref{3d-evolution} (m) where a 10 meV upward shift of $E_{\rm F}$ (electron-doped) breaks the connection among petals. As our system behaves as an electron-doped (as discussed earlier), we can easily assume that the four petals are disconnected from each other and electron move through the extremum orbit within a single petal as shown in FIG. \ref{3d-evolution} (m). The theoretically calculated extremum area of a single petal is very close to the experimental one which unambiguously supports the claim stated earlier. The DFT calculation slightly overestimates the FS area, indicating the electron-doped system. \\

The structure of the $\delta$ pockets in different angle are shown in FIG. \ref{3d-evolution} (c). Observing the 3D nature of the pockets we can say, area of the pocket in $k_{x}-k_{y}$ plane is much larger than the area along $k_{x}-k_{y}$ plane. Interestingly, $F_{\delta}$ from angle dependent dHvA changes according to this FS topology. The angular variation of $F_{\delta}$ is shown by the dashed magenta line in FIG. \ref{rotation}. The data suggest that when crystal lies in [001], $F_{\delta}$ has its highest value. But with the tilting towards [100],  $F_{\delta}$ started to decrease. As the area of the pocket in $k_{x}-k_{y}$ is very small, no signal was observed along [100] direction. Such an angle dependent study again strongly verify the nature of the FS.

\subsubsection {Frequency from $\alpha$-sheets ($F_{\alpha}$)} The frequency corresponding to the $\alpha$ sheet is not observed by us because the extremum area of the $\alpha$ and $\beta$ sheets are very close to each other (see TABLE-\ref{table-1}) and hence, both frequencies are superimposed to each other. We show how the FS evolves from hole-doping to electron-doping in FIG. \ref{3d-evolution} (g)-(i) for the completeness of our study. We observed that a 100 meV downward shift of $E_{\rm F}$ (hole doping) can break the $\alpha$ sheet into two equal parts. On the other hand 100 meV upward shift of $E_{\rm F}$ (electron doping)  enhance the volume of the FS. The extremum orbit in the $k_{x}-k_{y}$ plane is indicated by the red arrow in FIG. \ref{3d-evolution} (h).

\subsubsection{Frequency from $\gamma$-sheets ($F_{\gamma}$)} $F_{\gamma}$  is observed at 792 T in the FFT spectra as shown in FIG. \ref{fig:SdH} (a). The FS area corresponding to $\gamma$-sheets is the largest and most intense compared to all other pockets. The particular frequency comes from ellipsoid-shaped FS lying at the Z-point of the BZ. The 3D shape of the $\gamma$-pocket is shown in FIG. \ref{3d-evolution} (f)
The angle dependent study of $F_{\gamma}$ is shown by the blue dashed line in FIG. \ref{rotation}. It indicates that if we change the crystal orientation from [001] to [100], $F_{\gamma}$ changes from 790 T to 1124 T. The angular dependency of $F_{\gamma}$ suggests that the semi-major axis of the pocket is 1.5 times longer than its semi-minor axis. Such kind of experimental result is very consistent with our theoretically generated FS structure. Unlike the $\beta$ and $\delta$, experimental FS area of $\gamma$-pocket is slightly lower than the theoretically predicted one (see TABLE-\ref{table-1}). Lowering the FS area of any electron pocket is justifiable for an electron-doped system. The amplitude of $F_{\gamma}$ shows the adequate temperature dependency that help us to calculate the `effective mass'. All parameters\textit{ viz}. `effective mass' `Dingle temperature' and `Berry phase' derived from the  $F_{\gamma}$ are discussed in details in the subsequent section.
 
 \subsection{The study of effective mass, Dingle Temperature and Berry phase} 
 
\begin{table}[]
	\caption{Berry phase ($\phi_{B}$) calculated from $F_{\beta}$, $F_{\delta}$ and $F_{\gamma}$. $\phi_{B}$ of the corresponding frequencies are listed for various $\Delta$.}
\begin{tabular*}{0.49\textwidth}{@{\extracolsep{\fill} }  c  c  c  c  }

	\toprule\toprule
		\multirow{2}{*}{\begin{tabular}[c]{@{}c@{}}List of\\ frequencies\end{tabular}} &                                     & Berry phase ($\phi_{B}$)              &                   \\ 
		& $\Delta$ = -1/8  & $\Delta$ = 0      & $\Delta$ = 1/8    \\ 	\midrule
		
		$F_{\beta}$                                                                    & -0.5$\times \pi$ & -0.76$\times \pi$ & 1.01$\times \pi$  \\ 
		$F_{\delta}$                                                                   & 0.13$\times \pi$ & -0.13$\times \pi$ & -0.38$\times \pi$ \\ 
		$F_{\gamma}$                                                                   & 0.41$\times \pi$& 0.16$\times \pi$  & -0.06$\times \pi$ \\ \bottomrule\bottomrule
	\end{tabular*}
\label{berry-table}
\end{table}

\begin{table}[]
	
	\caption{Parameters derived from the $F_{\gamma}$ oscillation. $m^{\ast}$, effective mass; $T_D$, Dingle temperature; $\tau$, relaxation time and $\mu_q$, quantum mobility.}
	\begin{tabular*}{0.49\textwidth}{@{\extracolsep{\fill} }  c  c  c  c  }

		\toprule\toprule
	\multicolumn{1}{c}{\begin{tabular}[c]{@{}c@{}}$m^{\ast}$\\ (Units of $m_0$)\end{tabular}} & \begin{tabular}[c]{@{}c@{}}$T_D$\\ (K)\end{tabular} & \begin{tabular}[c]{@{}c@{}}$\tau$\\ (ps)\end{tabular} & \begin{tabular}[c]{@{}c@{}}$\mu_q$\\ ($m^2V^{-1}s{-1}$)\end{tabular} \\ \midrule
	0.40                                                                                     & 1.1                                                 & 1.2                                                   & 0.53                                                                 \\ \bottomrule\bottomrule
	\end{tabular*}
\label{dingle-table}

\end{table}

 \begin{figure*}
 	\includegraphics[width=0.98\textwidth]{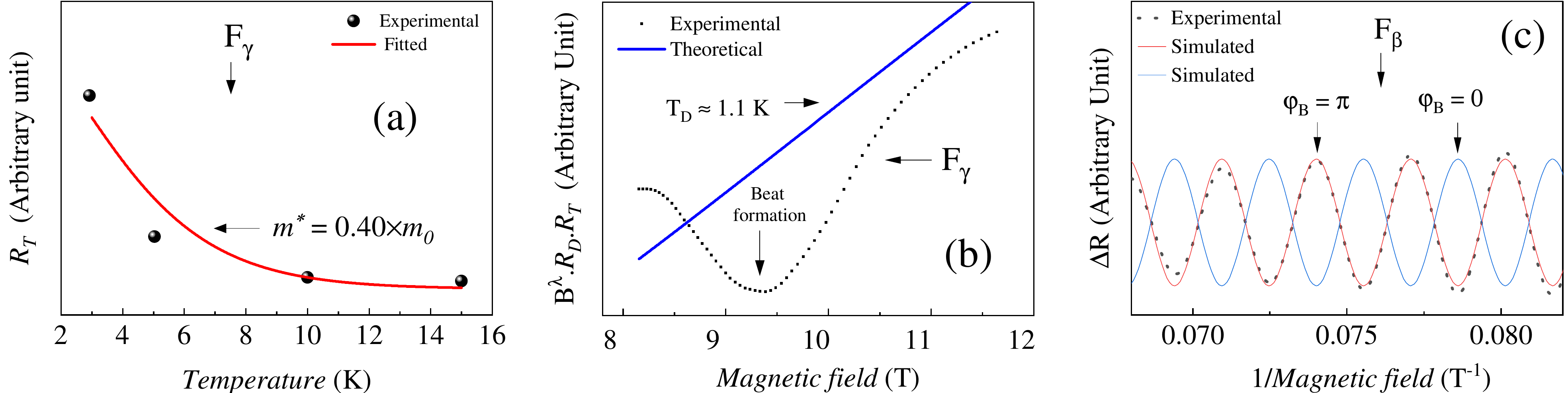}
 	\caption{(a) Calculation of the effective mass from \textit{T}-dependent amplitude of SdH oscillation. (b) The amplitude-simulation with varying $T_D$. (c) The phase part of the L-K formula is fitted for F$_{\beta}$. Simulated oscillations with varying $\phi_B$ are indicated by the red and blue line.}
 	\label{berry}
 \end{figure*}
The oscillation pattern of the resistivity is described by the Lifshitz-Kosevich (LK) formula
\begin{equation}
	\Delta R \propto -B^{\lambda}R_TR_{\rm D}R_S \sin \left[2\pi \left(\frac{F}{B}-\Gamma-\Delta\right)\right]
	\label{lk}
\end{equation}
where $R_T = \alpha T\mu/B\space \sinh(\alpha T\mu)$, $R_D = \exp(-\alpha T_D \mu/B)$ and $R_S = \cos(\pi g \mu /2)$. $\mu$ is the ratio of the effective cyclotron mass $m^{\ast}$ to free electron mass $m_0$. $T_{\rm D}$ is the Dingle temperature, and $\alpha$ = $(2\pi ^{2}k_Bm_0)/(\hbar e)$. The oscillation phase is described as a sine term with an additional phase factor $\Gamma-\Delta$, in which $\Gamma$ = $\frac{1}{2}$ - $\phi _B/2\pi$ and $\phi _B$ is the Berry phase. The dimensionality of FS is determined by the phase shift $\Delta$. For 2D FS, $\Delta$ = 0. For 3D FS, $\Delta$ can take value $\pm$ 1/8 according to the minima and maxima of the FS area. The term $\lambda$ is also determined by the FS dimensionality. $\lambda$ can takes value $ \frac{1}{2} $ and 0 for 3D and 2D FS respectively. From the L-K formula, effective mass  $m^{\ast}$ can be calculated from the fit of the thermal damping factor $R_T$ with the temperature dependence amplitude of the oscillation. From the SdH oscillation, we have observed that an adequate \textit{T}-dependent signal comes only from F$_{\gamma}$. From the \textit{T}-dependent amplitude-damping, we have calculated the effective mass ($m^{\star}$) of the electron. The fitting of R$_T$ for the F$_{\gamma}$ is shown in FIG. \ref{berry} (a) The derived m$^{\star}$ from the calculation is approximately 0.40$\times m_0$. Putting the value of $m^{\star}$, we have theoretically estimated the Dingle temperature ($T_{\rm D}$). The amplitude of the oscillation with varying $T_{\rm D}$ is shown in FIG. \ref{berry} (b). We have also marked the beat formation phenomena (discussed earlier) in the figure. The extraction of $T_{\rm D}$  from a beat forming data was also done earlier\cite{shoenberg2009magnetic}. Our estimated $T_{\rm D}$ from the fitted data is approximately 1.1 K.

The Dingle temperature of the compound has a microscopic origin closely related to the carrier's scattering. Our low  $T_{\rm D}$ indicates that sample has low imperfections \textit{viz.} dislocations and mosaic structure\cite{shoenberg2009magnetic}. We have fitted the phase part of the oscillation from the L-K formula for calculating the Berry phase. As all Fermi sheets of InBi have the maximum area, $\Delta$ of our case is equal to -1/8. For the completeness of our study, we have derived $\phi_B$ for $\Delta$ = 1/8, 0, and -1/8 for all the frequencies. Derived Berry phase for F$_{\beta}$, F$_{\delta}$ and F$_{\gamma}$  are listed in TABLE \ref{berry-table}. The estimated $\phi_B$ from F$_{\beta}$ is very close to $\pi$ whereas $\phi_B$ coming from F$_{\delta}$ and F$_{\gamma}$ are close to zero. The $\pi$-Berry phase of the band indicates the non-triviality of the FS. If the $\phi_B$ departure from $\pi$ to 0, the FS becomes non-trivial to trivial. Our study suggests except $\beta$-sheet, all other FS shows trivial nature. Using Dingle temperature $T_{\rm D}$ ($\sim$ 1.1 K), we have calculated the quantum relaxation time $\tau_q = \hbar/ 2\pi k_{\rm B}T_{\rm D} = 1.20\times10^{-12} s$ and quantum mobility $\mu_q = e\tau_q/m^{\star} = 0.53 m^{2}V^{-1}s^{-1}$. All estimated parameters from the F$_{\gamma}$ are listed in TABLE \ref{dingle-table}. The quantum mobility ($\mu_q$) at 3 K is very close to the compound's effective mobility ($\mu_{eff}$) estimated from Hall data at the same temperature. The consistent result suggests that our experimental outcome from SdH oscillation and the Hall measurement is very reliable. Our estimated quantum mobility is slightly lower than the classical mobility. Such a phenomenon is expected because quantum mobility is sensitive to both small-angle and large-angle scattering. On the other hand, classical Drude's mobility is only influenced by large-angle scattering\cite{narayanan2015linear}.

\section{Conclusion}

We presented a detailed discussion of transport phenomena and the Fermi surface topology of InBi. The turn-on phenomena of resistivity are intensely studied. We gave an in-depth analysis of the origin of turn-on behavior that justified our mathematical model. We have calculated the temperature-dependent carrier density and mobility for both carriers. We observed electron-hole carrier compensation follows up to 25 K. The high MR below 25 K indicates carrier compensation is one of the primary origins of the XMR. We successfully resolved the 3D structure of every electron and hole pocket with the help of Shubnikov-de Haas oscillation. Our study reveals that two bands, each electron and hole-like, majorly contribute to transport phenomena. The holes contribute from one pocket lying at the central and four pockets lying at the corner of the BZ. Similarly, the electrons come from one pocket lying at the body center and four pockets at the edge of the BZ. The collective participation of each electron and hole pockets makes the compound nearly perfect compensated semi-metal at low temperature. The modulation of electron and hole pockets occurred as if the compound behaved slightly electron doped at 3 K. The structure of each Fermi sheets excellently matches with the theoretical result. The unique Fermi surface topology governing the carrier compensation is closely correlated to the magneto-transport phenomena of the compound.

\begin{acknowledgments}
SP acknowledges the financial support from DST-SERB, India under the project no. ECR/2017/001243. One of the authors Shovan Dan would like to thank SERB, India (Grant No. PDF/2021/004624) for the financial assistance. We thank Dr. Arkadeb Pal for helping in a scientific collaboration.
 \end{acknowledgments}

\nocite{*}

\bibliography{ref}

\end{document}